# Role of Jahn-Teller disorder in Raman scattering of mixed-valence manganites

M. N. Iliev,[1] M. V. Abrashev,[2] V. N. Popov,[2] and V. G. Hadjiev[1]
[1]*Texas Center for Superconductivity and Advanced Materials, University of Houston, Houston, Texas 77204-5002, USA*
[2]*Faculty of Physics, University of Sofia, 1164 Sofia, Bulgaria*

The mixed-valence perovskitelike manganites are characterized by the unique interrelation of Jahn-Teller distortions, electric and magnetic properties. The Jahn-Teller distortion follows the $Mn^{3+} \rightarrow Mn^{4+}$ charge transfer with some delay. Its development depends on the lifetime of Mn in the $3+$ state, governed by the $Mn^{4+}/Mn^{3+}$ ratio and magnetic correlations. The noncoherence of Jahn-Teller distortions in orthorhombic mixed-valence manganites and rhombohedral $RMnO_3$ ($R$ = rare earth) results in oxygen disorder. We demonstrate that the Raman spectra in this case are dominated by disorder-induced bands, reflecting the oxygen partial phonon density of states (PDOS). The PDOS origin of the main Raman bands in insulating phases of such compounds is evidenced by the similar line shape of experimental spectra and calculated smeared PDOS and disappearance of the PDOS bands in the ordered ferromagnetic metallic phase.

The structure of the perovskitelike manganites $A_{1-x}A'_x MnO_3$ ($A$ = trivalent rare earth, $A'$ = divalent alkaline earth) differs from that of an ideal cubic perovskite by mainly in-phase or out-of-phase rotations of $MnO_6$ octahedra around some of the cubic axes[1] and stretching distortions of $Mn^{3+}O_6$ (but not $Mn^{4+}O_6$) octahedra due to the Jahn-Teller (JT) effect. At a microscopic level the structure has exact translational symmetry in only few cases, such as the orthorhombic phase (*Pnma*) of $LaMnO_3$ and $CaMnO_3$ or the charge- and orbital-ordered (COO) phase ($P2_1/m$) of $La_{0.5}Ca_{0.5}MnO_3$. In the general case, the real local structure of mixed-valence $A_{1-x}A'_x MnO_3$ *a priori* differs from the averaged one (as determined by diffraction studies) due to the random distribution of $A^{3+}$ ($A'^{2+}$) and $Mn^{3+}$ ($Mn^{4+}$) ions, respectively. The translational symmetry in this case is only an approximation, and the effects of atomic disorder have to be accounted for.

The random distribution of $Mn^{3+}$ and $Mn^{4+}$ at the $B$ sites determines to a great extent the electronic transport and magnetic ordering. The $Mn^{3+}/Mn^{4+}$ charge and orbital disorder results in structural disorder, too. The latter disorder is mainly due to the difference of the bond lengths in noncoherently distributed $Mn^{3+}O_6$ and $Mn^{4+}O_6$ octahedra. Its existence, including in the metallic phase, has been confirmed experimentally in x-ray,[2,3] neutron scattering,[2,4,5] and photoemission[6] experiments on $La_{1-x}Ca_x MnO_3$ and related $La_{1.2}Sr_{1.8}Mn_2O_7$.

The disorder in the doped rare-earth manganites varies strongly with doping level and temperature. Its dynamics is directly related to such electronic processes as charge hopping and localization in the insulating paramagnetic phase and charge delocalization in the metallic ferromagnetic phase. One important issue, concerning the metallic phase, is whether it is homogeneous or coexists with another (possibly disordered) insulating phase both above and below $T_c$, as supposed by some theoretical models.[7] Therefore, monitoring of the disorder is of significant interest for understanding the interplay of transport, magnetic, and structural properties.

The Raman spectroscopy is an efficient tool for the study of structural disorder, including the dynamical one. The first-order Raman phonon spectrum of a nominally perfect crystal consists of series of narrow lines, which correspond to Raman-allowed zone-center ($\Gamma$-point) phonon modes and obey definite polarization selection rules. The Raman scattering selection rules imposed by the crystal symmetry do not exist in the opposite limit case of amorphous material. The Raman spectrum in this case roughly reflects the one-phonon density of states of the crystalline form of the same material.[8,9] The doped rare-earth manganites are an example of partial structural disorder. While the $A$(La/Ca) and $B$(Mn) sublattices in a good approximation keep their translational symmetry, the O(oxygen) sublattice is strongly distorted. Compared to the Raman spectra of undistorted material one may expect such effects of partial disorder: (i) broadening of some or all first-order Raman lines, (ii) activation in the Raman spectrum of otherwise forbidden phonon modes, and (iii) the appearance of additional broad Raman bands of phonon density-of-states origin. It is plausible to assume that the main contribution to the density-of-states spectral features will come from the most strongly distorted oxygen sublattice.

In recent years there have been numerous reports on the variations of the Raman spectra of $R^{3+}_{1-x}A^{2+}_x MnO_3$ ($R$ = rare earth, $A$ = Ca,Sr,Ba, $0 \leq x \leq 1$) with $x$ and temperature.[10–16] As a rule, the Raman spectra of doped manganites consist of two or three broad bands at positions close to those of the strongest Raman lines for the parent orthorhombic $RMnO_3$ compounds with *Pnma* structure. On the basis of this closeness, the broad bands in the Raman spectra of doped manganites have usually been assigned to the corresponding Raman modes in the *Pnma* structure.[14] The width of these bands, however, remains practically unchanged in a wide temperature range and is too large, which rules out an explanation by thermal broadening. Alternatively, the broad bands have also been assigned to two-phonon scattering[10] or activation of otherwise forbidden Raman modes.

In this paper we analyze the Raman spectra of doped manganites in order and disorder terms. On the example of $La_{1-x}Ca_x MnO_3$ ($0 \leq x \leq 1$) in close comparison with other

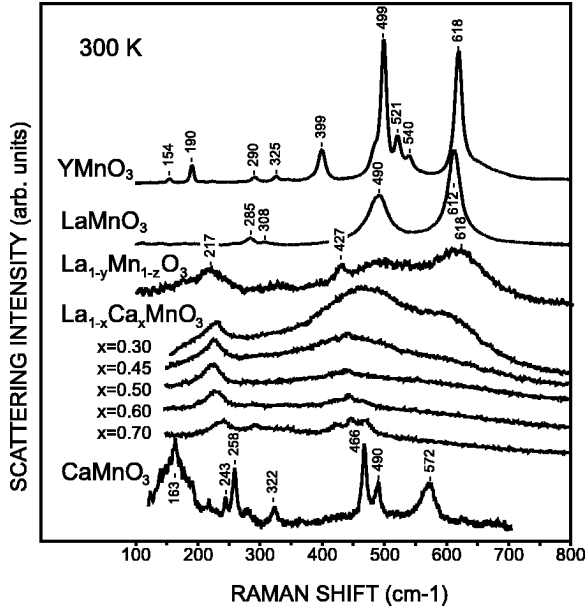

FIG. 1. Comparison of the Raman spectra of $YMnO_3$, $LaMnO_3$, $La_{1-y}Mn_{1-z}O_3$, $La_{1-x}Ca_xMnO_3$ ($x=0.3, 0.45, 0.5, 0.6, 0.7$), and $CaMnO_3$ at room temperature

related materials we argue that the Raman spectra reflect in a unique way the structural disorder induced by noncoherent Jahn-Teller distortions, as well as its variations with $x$ and temperature.

The samples and the experimental setups used to obtain the experimental spectra of $YMnO_3$ (*Pnma*), $LaMnO_3$ (*Pnma*), $LaMnO_3$ ($R\bar{3}c$), $La_{1-x}Ca_xMnO_3$, and $CaMnO_3$ (*Pnma*) are described elsewhere.[16–20] The calculation of the PDOS and partial PDOS was performed using a shell model of the lattice dynamics developed for $LaMnO_3$.[17] The sampling was carried out over 30 000 points in the irreducible wedge of the orthorhombic or rhombohedral Brillouin zone. In the case of the partial PDOS, only modes with predominant oxygen participation were considered.

$LaMnO_3$ contains only $Mn^{3+}$ ions. The JT distortions are compatible with the orthorhombic (*Pnma*) structure and therefore no disorder effects are expected for stoichiometric samples. The most intensive Raman lines at 612 cm$^{-1}$ ($B_{2g}$), ~490 cm$^{-1}$ ($A_g$), and ~285 cm$^{-1}$ ($A_g$) (see Fig. 1) correspond to stretching, bending, and rotational oxygen vibrations, respectively.[17] These lines are broader than in the isostructural $YMnO_3$, and their width and position vary in the spectra reported by different groups.[10,14–16] This indicates a larger deviation of the real structure of nominally stoichiometric $LaMnO_3$ from the ideal *Pnma* structure, presumably due to some oxygen deficiency.[21]

As illustrated in Fig. 1, the substitution of Ca for La changes drastically the Raman spectra of $La_{1-x}Ca_xMnO_3$. For $x<0.5$ the spectra in the paramagnetic phase are characterized by two rather broad (~100 cm$^{-1}$) high-frequency bands centered at ~500 and ~600 cm$^{-1}$ and a narrower (15–20 cm$^{-1}$) low-frequency peak. Remarkably, such a line shape of the Raman spectra is characteristic not only for $La_{1-x}Ca_xMnO_3$, which remains orthorhombic over the whole $0 \leq x < 0.5$ range. Similar spectra exhibits $La_{1-x}Sr_xMnO_3$, which undergoes an orthorhombic-to-rhombohedral transition at $x \approx 0.1$, as well as nominally undoped rhombohedral $LaMnO_3$ (Ref. 16 and 18) and cation-deficient $La_{1-y}Mn_{1-z}O_3$ (Ref. 16). What these materials have in common is the disorder of the oxygen sublattice due to either the coexistence of $Mn^{3+}$ and $Mn^{4+}$ or noncompatibility of Jahn-Teller distortions with the rhombohedral structure. As the transformation of the relatively sharp Raman lines (corresponding to oxygen $\Gamma$-point vibrations) to broadbands correlates with the occurrence of oxygen disorder, it is reasonable to assign the latter bands to disorder-induced scattering. A similar disorder effect occurs in nominally stoichiometric orthorhombic $LaMnO_3$ at high temperatures. Indeed, the Raman spectra of $LaMnO_3$ at 520 and 720 K (Ref. 22), are almost identical to these of $La_{0.7}Ca_{0.3}MnO_3$.

It is generally accepted that the spectral line shape of disorder-induced Raman scattering qualitatively reflects the phonon density of states (PDOS) versus phonon energy distribution for the perfect crystal. Shuker and Gammon[8] have shown that the Raman spectrum $I(\omega)$ of a disordered material can be approximated by

$$I(\omega) = \sum_b C_b [n(\omega,T)+1] \rho_b(\omega) \omega^{-1}, \quad (1)$$

where $\rho_b(\omega)$ is the density of vibrational states in band $b$ and $n(\omega,T)$ is the Bose factor. The factors $C_b$ contain the polarization dependence of scattering and allow different vibrational bands to couple with varying strength to the radiation. If all bands couple equally to the incident light $C_b = C$, the reduced Raman spectrum $I'(\omega) = I(\omega)\omega[n(\omega,T)+1]^{-1} = C\rho(\omega)$ is directly related to the total phonon density of states, $\rho(\omega)$.

A comparison of the partial (oxygen) PDOS profiles for orthorhombic and rhombohedral $LaMnO_3$, as obtained from lattice dynamical calculations (Fig. 2, top panels), with the Raman spectra of $La_{0.7}Ca_{0.3}MnO_3$ (*Pnma*) and $LaMnO_3(R\bar{3}c)$ (Fig. 2, bottom panels), assumed to be dominated by disorder scattering features, shows that the experimentally observed bands are much broader than the calculated widths of the phonon branches, involving oxygen motions. The width of the PDOS maxima, however, can be fitted to that of the Raman bands by replacing the PDOS $\rho(\omega)$ by a smeared PDOS,

$$\rho_s(\omega) = \int_{-\infty}^{\infty} S(\gamma, \omega - \omega') \rho(\omega') d\omega', \quad (2)$$

and varying the width $\gamma$ of the smearing function $S(\gamma, \omega - \omega')$. The smearing of the vibrational frequency is due to a finite vibrational time and therefore it is reasonable to use as a smearing function the Lorentzian

$$S(\gamma, \omega - \omega') = \frac{1}{\pi} \frac{\tau}{1 + \tau^2(\omega - \omega')^2}, \quad (3)$$

with half-width $\gamma = 1/\tau$.

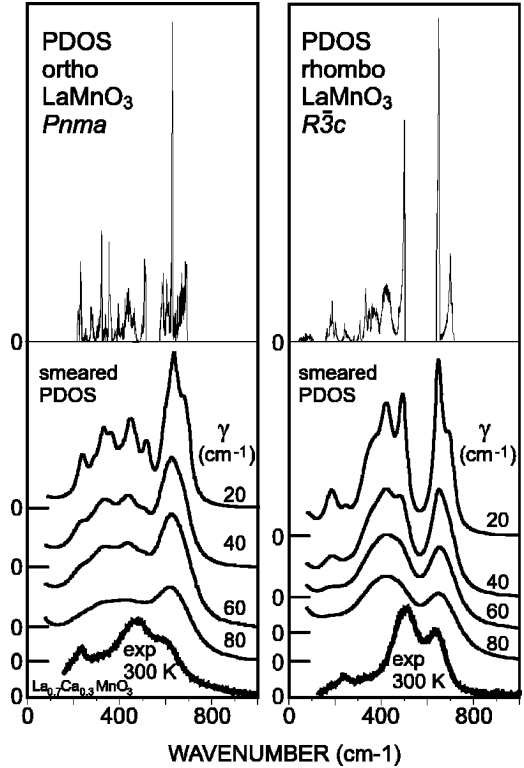

FIG. 2. Comparison of partial PDOS (top panels) and smeared partial PDOS for $\gamma=20, 40, 60,$ and $80$ cm$^{-1}$ (middle panels) of orthorhombic (left) and rhombohedral (right) LaMnO$_3$ with experimental Raman spectra of orthorhombic La$_{0.7}$Ca$_{0.3}$MnO$_3$ and rhombohedral LaMnO$_3$.

The smeared PDOS profiles, shown in Fig. 2, correspond to the smeared partial PDOS functions calculated with $\gamma = 20, 40, 60,$ and $80$ cm$^{-1}$, respectively. For $\gamma=40$ cm$^{-1}$ there is good correspondence between the position and width of the main maxima of PDOS and the experimental Raman spectra. The difference in relative intensities of the maxima of the so-calculated PDOS and Raman spectra as well as some shift of the maxima is not unexpected for at least two reasons: (1) the electron-phonon interaction is not the same for the different phonon branches and (2) there are some differences in the phonon structure of orthorhombic LaMnO$_3$ and La$_{0.7}$Ca$_{0.3}$MnO$_3$. The parameter $\tau$ has the physical meaning of the mean phonon lifetime. It is determined by the scattering rate from (quasi)static lattice distortions ($1/\tau_d$), other phonons ($1/\tau_{anh}$), and the time between two consequent hops ($\tau_h$) as the acts of scattering and atomic rearrangement interrupt the coherency of atomic vibrations. In the case of static or quasistatic disorder $\tau_d \ll \tau_h$ and $\tau_d \ll \tau_{anh}$. The latter relation is justified by the much smaller Raman linewidths for ordered orthorhombic YMnO$_3$, LaMnO$_3$, and CaMnO$_3$, where the broadening is determined by mainly $\tau_{anh}$. Therefore,

$$\gamma = \frac{1}{\tau} \approx \frac{1}{\tau_d} \qquad (4)$$

is governed mainly by the lattice distortions.

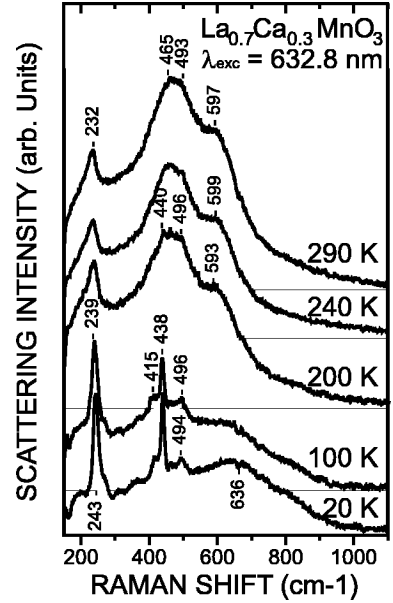

FIG. 3. Variation with temperature of the Raman spectra of La$_{0.7}$Ca$_{0.3}$MnO$_3$ ($T_c = 200$ K).

It is reasonable to expect that the intensity of disorder-induced Raman structures in rare-earth manganites will depend on both the doping level and temperature. Doping increases the relative volume of distorted areas, thus the contributions to the Raman spectra of spectral structures of PDOS origin. For the orthorhombic (*Pnma*) materials the increase with $x$ of the distorted volume and PDOS bands is accompanied by a decrease in intensity and broadening of the sharp $\Gamma$-point Raman lines. The distorted volume reaches its maximum for disordered Mn$^{3+}$/Mn$^{4+}$ composition with $\sim 30\%$ Mn$^{4+}$. As pointed out by Goodenough,[23] this is the disordered composition with the largest number of Mn$^{3+}$ ions with one and only one Mn$^{4+}$ nearest neighbor. Figure 1 demonstrates that in the insulating phase, the intensity of the density-of-states bands does decreases for $x>0.3$. The decrease in intensity of the disorder-induced bands at higher doping levels and the appearance of some sharp Raman features can be explained by increasing the volume of domains with short-range COO (Ref. 19).

The single act of Mn$^{+3} \rightarrow$ Mn$^{+4}$ hop is much faster than the time of atomic rearrangement caused by the charge transfer. The Jahn-Teller distortion develops (at Mn$^{3+}$) or disappears (at Mn$^{4+}$) after some delay time $\tau_{JT}$, which can reasonably be approximated by the period of Jahn-Teller-like phonon vibration, $\approx 10^{-13}$–$10^{-14}$ sec. Given that the mean lifetime $\tau_h$ of Mn in the 3+ state between two subsequent Mn$^{+3} \rightarrow$ Mn$^{+4}$ hops remains much larger than $\tau_{JT}$, the dynamical Jahn-Teller distortions can be treated as quasistatic with respect to the Raman scattering processes. Therefore, the broadening of the Jahn-Teller-disorder-induced Raman bands, if observed, will be practically temperature independent, provided

$$\tau_d > \tau_h \gg \tau_{JT} \approx 10^{-13}\text{--}10^{-14} \text{ sec}, \qquad (5)$$

and the Jahn-Teller distortions are not coherent. For the doped rare-earth manganites this is fulfilled in the nonmetallic paramagnetic, antiferromagnetic, and ferromagnetic phases at doping levels $x<0.5$. Experimental confirmation of such kind of behavior can be found in Refs. 11 and 13–15.

In the metallic state $\tau_h$ strongly decreases and the relation (5) is no longer valid. Indeed, at $\tau_h \lesssim \tau_{JT}$ the Jahn-Teller distortion cannot develop or only partly develops in the time frame of the Mn$^{3+}$ state. Physically, this results in a strong reduction or disappearance of the Jahn-Teller effect and related noncoherent lattice distortions. This disorder-order transition, concomitant with the paramagnetic-to-ferromagnetic and insulator-metal transitions, is reflected in the Raman spectra by the reduction or disappearance of the broad Jahn-Teller-distortion-induced bands and the appearance of much sharper Raman lines, corresponding to the $\Gamma$-point Raman phonons of the ordered structure. Figure 3 illustrates such kind of behavior in the example of La$_{0.7}$Ca$_{0.3}$MnO$_3$ ($T_c=200$ K). The remnants of broad Jahn-Teller bands well below $T_c$ indicate that some oxygen disorder is still present in the metallic state. This result is of definite importance as it supports the model of coexisting metallic and insulating domains and percolative mechanism of conductivity.[7]

Our results demonstrate that Raman spectroscopy can efficiently be applied for monitoring Jahn-Teller disorder in mixed-valence manganites, which opens a new way to control the real structure of CMR materials at a microscopic level.

This work was supported in part by the state of Texas through the Texas Center for Superconductivity and Advanced Materials.